\documentclass[sigconf]{acmart}

\AtBeginDocument{%
  \providecommand\BibTeX{{%
    \normalfont B\kern-0.5em{\scshape i\kern-0.25em b}\kern-0.8em\TeX}}}

\copyrightyear{2022}
\acmYear{2022}
\setcopyright{rightsretained} 

\acmConference[MM '22]{Proceedings of the 30th ACM International Conference on Multimedia}{October 10--14, 2022}{Lisboa, Portugal}
\acmBooktitle{Proceedings of the 30th ACM International Conference on Multimedia (MM '22), Oct. 10--14, 2022, Lisboa, Portugal}
\acmISBN{978-1-4503-9203-7/22/10}
\acmDOI{10.1145/3503161.3551586}

\usepackage{etoolbox}
\makeatletter
\patchcmd{\maketitle}{\@copyrightpermission}{
   \begin{minipage}{0.3\columnwidth}
     \href{https://creativecommons.org/licenses/by-nc/4.0/}{\includegraphics[width=0.90\textwidth]{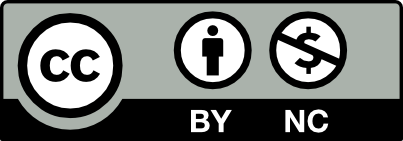}}
   \end{minipage}\hfill
   \begin{minipage}{0.7\columnwidth}
     \href{https://creativecommons.org/licenses/by-nc/4.0/}{This work is licensed under a Creative Commons Attribution-NonCommercial International 4.0 License.}
   \end{minipage}

   \vspace{5pt}
}{}{}

\makeatother

\begin{document}

\title {Deep Learning-Based Acoustic Mosquito Detection in Noisy Conditions Using Trainable Kernels and Augmentations}

\author{Sean Campos}
\authornote{Both authors contributed equally to this research.}
\email{sean.campos@berkeley.edu}
\orcid{0000-0001-8067-909X}
\author{Devesh Khandelwal}
\authornotemark[1]
\email{deveshkhandelwal@berkeley.edu}
\orcid{0000-0002-5875-998X}
\affiliation{%
  \institution{School of Information}
  \institution{University of California, Berkeley}
  \streetaddress{P.O. Box 1212}
  \city{Berkeley}
  \state{California}
  \country{USA}
  }

\author{Shwetha C. Nagaraj }
\email{shwethacn@ischool.berkeley.edu}
\affiliation{%
\institution{School of Information}
  \institution{University of California, Berkeley}
  \streetaddress{P.O. Box 1212}
  \city{Berkeley}
  \state{California}
  \country{USA}
}
  
\author{Fred Nugen}
\email{nooj@berkeley.edu}
\orcid{0000-0002-6761-7035}
\affiliation{%
  \institution{School of Information}
  \institution{Division of Computing, Data Science, and Society}
  \institution{University of California, Berkeley}
  \streetaddress{P.O. Box 1212}
  \city{Berkeley}
  \state{California}
  \country{USA}
  \postcode{43017-6221}
}

\author{Alberto Todeschini}
\email{todeschini@berkeley.edu}
\affiliation{%
  \institution{School of Information}
  \institution{Division of Computing, Data Science, and Society}
  \institution{University of California, Berkeley}
  \streetaddress{P.O. Box 1212}
  \city{Berkeley}
  \state{California}
  \country{USA}
  \postcode{43017-6221}
}
\renewcommand{\shortauthors}{Sean Campos et al.}

\begin{abstract}
In this paper, we demonstrate a unique recipe to enhance the effectiveness of audio machine learning approaches \cite{camastra2015machine} by fusing pre-processing techniques into a deep learning model \cite{lecun2015deep}. Our solution accelerates training and inference performance by optimizing hyper-parameters through training instead of costly random searches to build a reliable mosquito detector from audio signals. 
The experiments and the results presented here are part of the MOS-C submission of the ACM'22 challenge \cite{Schuller22-TI2}.  Our results outperform the published baseline by 212\% on the unpublished test set. We believe that this is one of the best real-world examples of building a robust bio-acoustic system that provides reliable mosquito detection in noisy conditions.

\end{abstract}

\ccsdesc[100]{Computer systems organization}
\ccsdesc[500]{Computing methodologies~Simulation evaluation}
\ccsdesc[500]{Computing methodologies~Computer vision representations}

\keywords{neural networks, bio acoustics, mosquito event detection}



\begin{teaserfigure}
  \includegraphics[width=\textwidth]{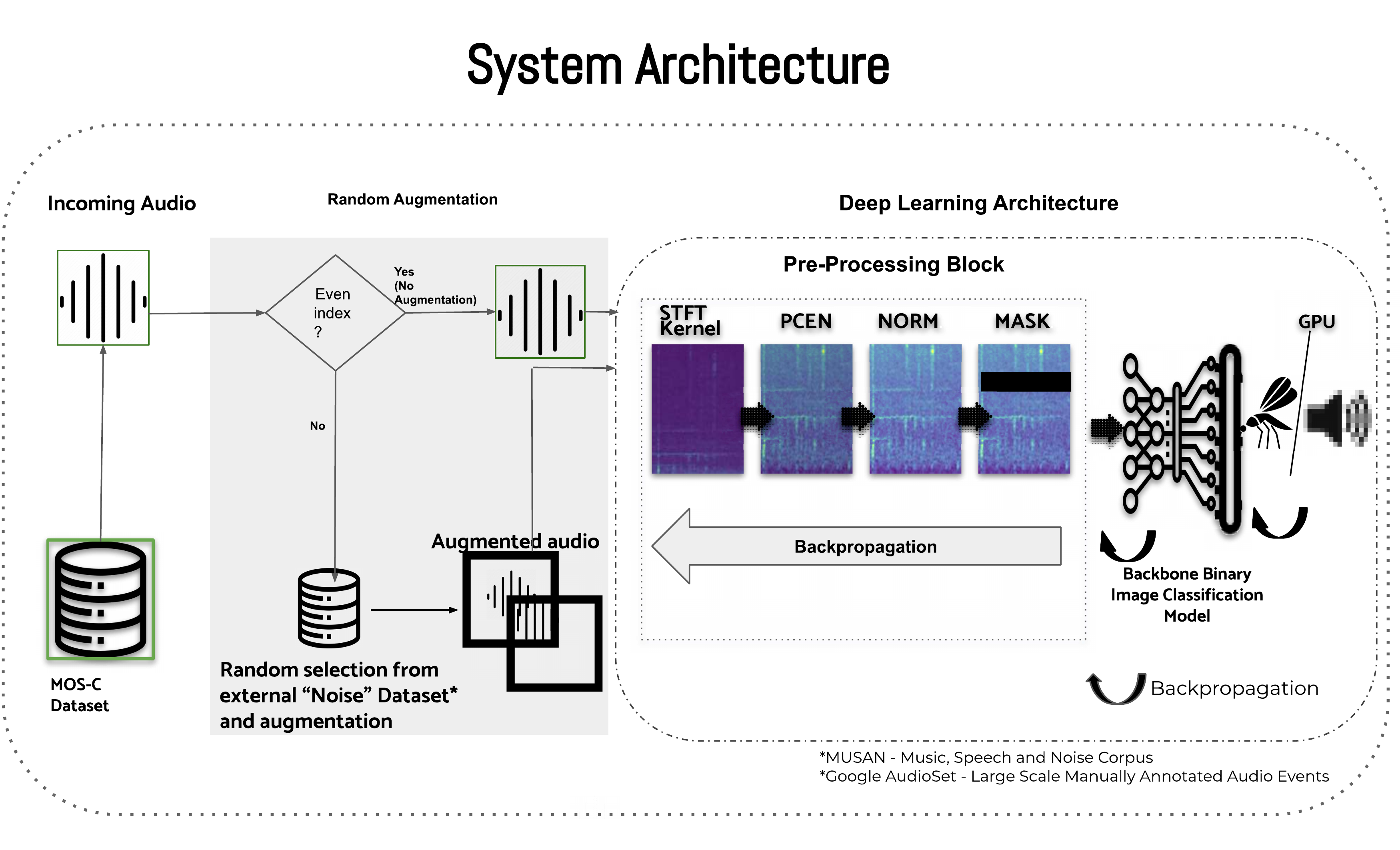}
  \vspace{-3em}
  \caption{A High-level Overview of the VecNet Mosquito Detection System}
  \Description{We use a unique combination of a deep-learning model (ConvNeXt) and data pre-processing techniques like PCEN, trainable kernels, and external data augmentation to create a model to detect mosquitoes.}
  \label{fig:Sys_Arch}
  \end{teaserfigure}
\maketitle

\section{Introduction}
Mosquitoes kill one million people per year \cite{manguin2013anopheles} and are the deadliest animal to humans on the planet. Global warming has increased the climatic suitability of mosquitoes in already endemic areas and about 1.4 billion additional people are at risk of malaria and dengue in urban areas in Africa and Southeast Asia alone. Due to Covid-19, critical 2020 milestones of The World Health Organization’s global malaria strategy have been missed, and without immediate action, the 2030 targets will not be met \cite{monroe2022reflections}. 
There is an urgent need and an incredible opportunity amongst entomologists, researchers, and health organizations around the world, to use innovative approaches in the identification of mosquitoes at scale, which will, in turn, improve vector control and intervention strategies.
Researchers have been exploring novel ideas to leverage budget smartphones as acoustic sensors \cite{kiskin2021humbugdb}. Mobile services are expanding in sub-Saharan Africa \cite{kshetri2022economics} and these have the potential to become affordable, non-invasive, automated real-time mosquito monitoring tools that can be deployed in homes at no extra cost. The approach will help in producing much-needed surveillance data of mosquito species behavior and distribution over extensive temporal and spatial scales to assess ongoing vector-control measures. This AI-driven solution can replace time-consuming and expensive manual survey and classification tasks (e.g. Polymerase Chain Reaction (PCR) tests \cite{snounou1993identification}) in remote and resource-constrained areas that bear the brunt of mosquito-borne diseases \cite{lenton2008tipping}.

\section{Data}
We use the publicly available dataset released as part of the ComParE \href{https://doi.org/10.48550/arxiv.2205.06799}{challenge}. The challenge dataset can be downloaded from \href{https://zenodo.org/record/6478589}{Zenodo}. Its attributes and the collection methods are described in detail in the Humbug system \cite{kiskin2021humbugdb}, while the splits and evaluation protocols are described as part of the challenge \cite{Schuller22-TI2}. 
To validate model performance the organizers provide two development sets:
\begin{itemize}
\item Dev A: Recordings from Tanzania containing mosquito audio obtained from phones placed in bednets.
\item Dev B: Recordings from the UK containing mosquito audio from lab-cultured mosquito larvae.
\end{itemize}
The baseline scores on the Dev sets are published as part of the challenge \cite{Schuller22-TI2}. As evident from the scores, Dev B, which features a lower signal-to-noise ratio (SNR) \cite{johnson2006signal}, is a more challenging subset.

\section{Our Approach}
We use a combination of a deep-learning model and data pre\-pro\-cessing techniques like Per-Channel Energy Normalization (PCEN), trainable kernels, and external data augmentation to create a model to detect mosquitoes. In the sections below we describe our unique data processing pipeline. A reproducible pipeline is available on our GitHub.\footnote{\url{https://github.com/seancampos/ComParE2022_VecNet/}}

\subsection{Pre-Processing}
To extract useful features from the audio samples, we convert them to their time-frequency representation: a spectrogram. This is done by applying the Short Time Fourier Transform (STFT) \cite{durak2003short} operation to the audio signals. Converting an audio sample into its time-frequency representation is a common practice in extracting patterns from raw audio \cite{altes1980detection}. The spectrograms are then passed as input to a computer vision backbone \cite{shah2022music} for further processing.
Figure 2 below shows different representations of audio. The top portion represents the raw form (a time series), while the bottom part represents its time-frequency equivalent after applying STFT.

\begin{figure}[h]
  \centering
    \includegraphics[width=\linewidth]{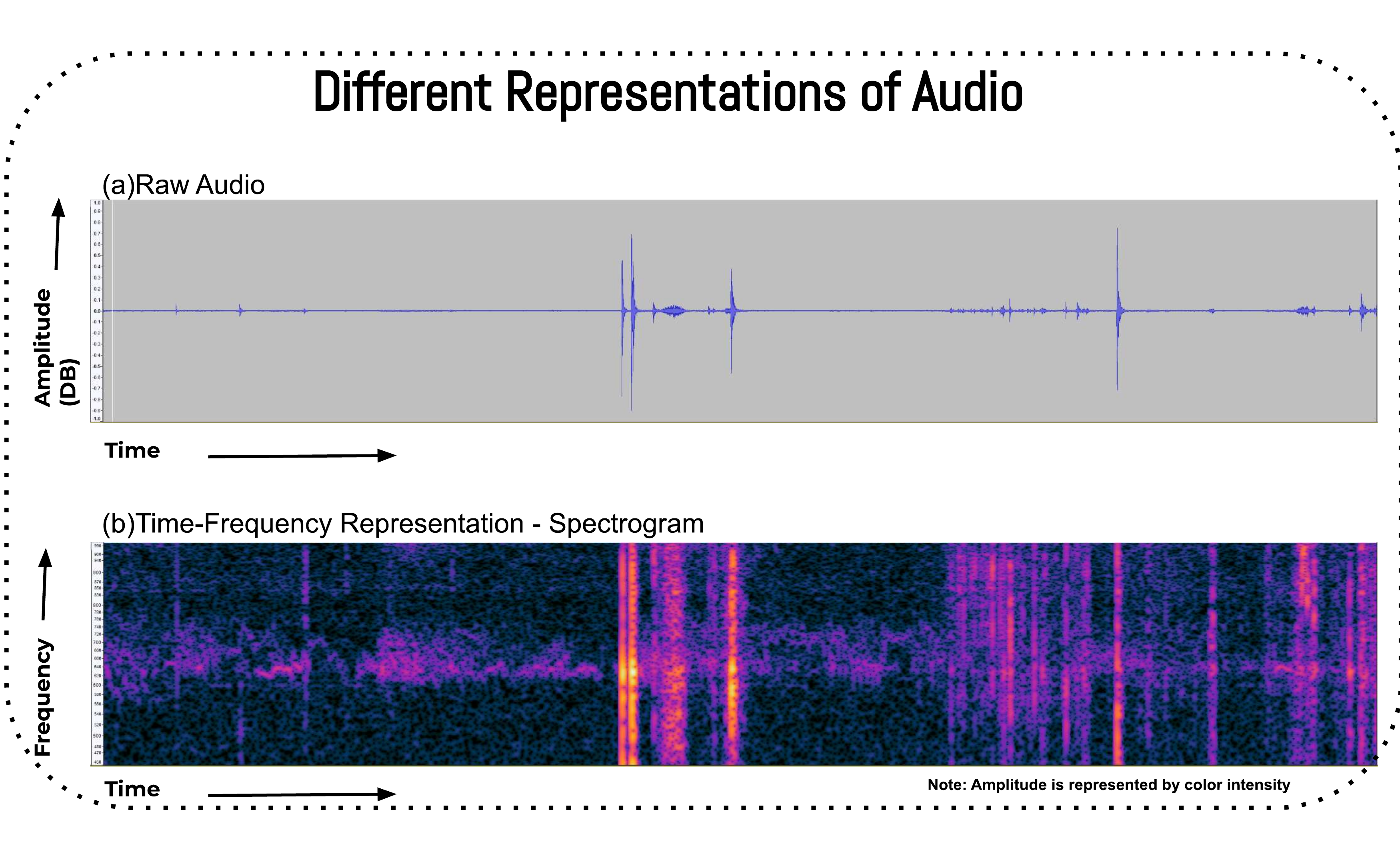}
  \caption{Waveform and Time-Frequency Representation of an Audio Signal}
  \Description{A sample spectrogram from raw audio.}
  \end{figure}

The audio samples in the dataset exhibit wide variations (owning to different recording conditions and the presence of background noise) and as a result, there is a wide variety of spectrograms that are sent as input to our backbone architecture. As an example consider Figure 3, which shows spectrogram representations of three randomly selected audio files in the dataset. The leftmost part shows a signal with uniform mosquito activity throughout its duration. The middle part shows a signal with sporadic instances of the presence of mosquito while the rightmost portion shows less noise and a clear pattern in fundamental mosquito frequency and its harmonics. The task to extract a discernible pattern gets even more complex due to the various hyperparameter settings of the STFT operation. 

\begin{figure}[h]
  \centering
    \includegraphics[width=\linewidth]{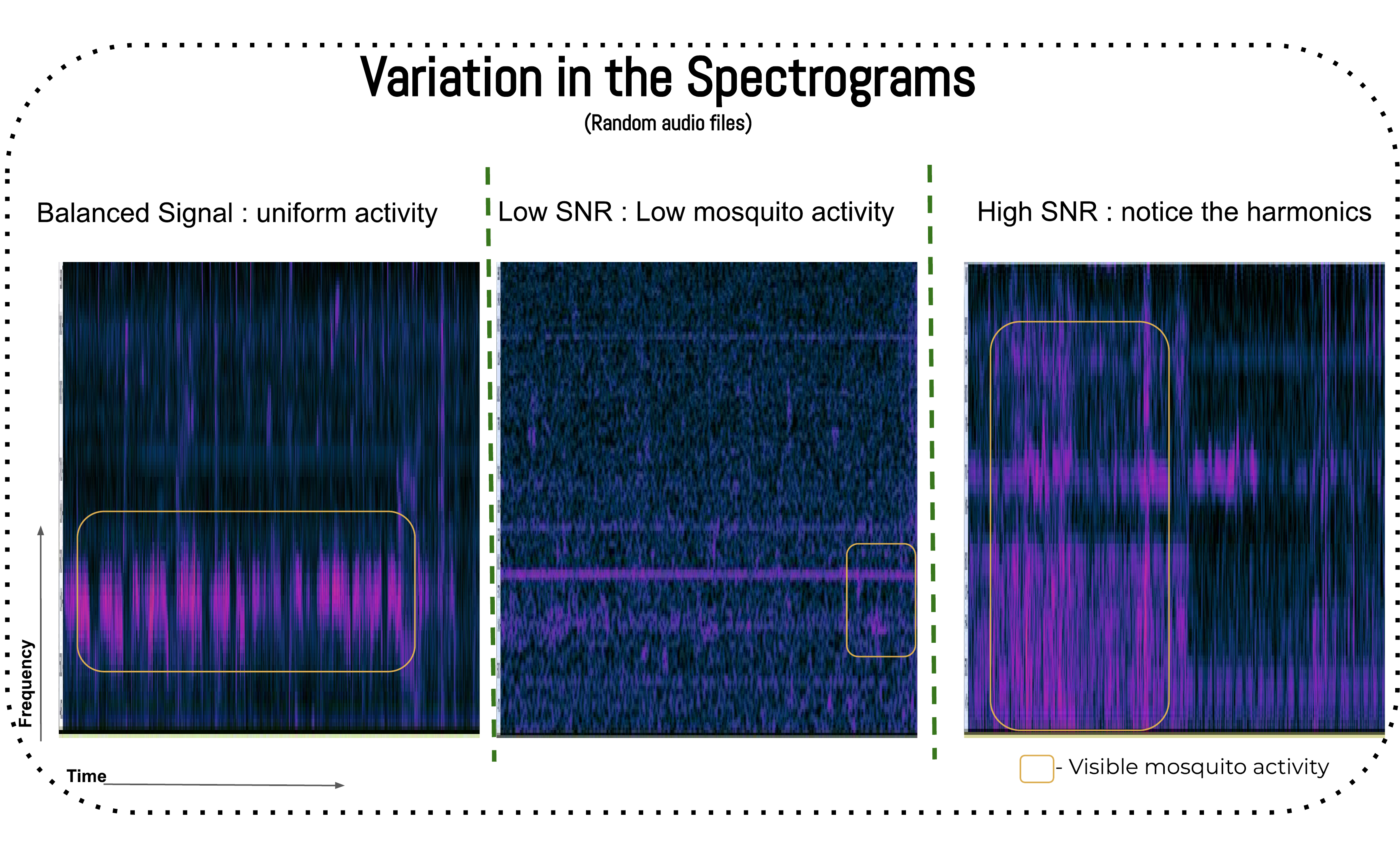}
  \vspace{-2em}
  \caption{Spectrogram Representation of a Few Randomly Selected Files}
  \Description{Wide Range of Spectrograms.}
  \end{figure}

\subsection{Augmentation}
We introduce random augmentations on the audio samples so that the model can learn to separate mosquito buzz from a variety of soundscapes. During training, for every other recording, we randomly sample files from an external database and add their corresponding tensor value to the tensor representation of the incoming audio (as indicated in Fig 1). The primary source of external sound files is the "noise" class in MUSAN \cite{snyder2015musan}, which is a general library of noise, speech, and music.  
We also supplement the raw files in the dataset with selected categories from Google Audioset \cite{gemmeke2017audio} based on specific misclassifications in our model, including vehicles, other insects, and babies crying.
Unlike the baseline solution, we do not discard any audio file owing to its duration. For a short audio file (less than 1.92 seconds), we either pad it with silence or augment it with audio from the two datasets, MUSAN and Audioset. Our training and inference sets consist of 1.92-second audio files.
Please refer to our \href{https://github.com/seancampos/ComParE2022_VecNet/}{repository} for examples.
While random augmentations might impact reproducibility, but it adds robustness to the model and makes it more generic

\subsection{Trainable Front-end}
In deep learning based acoustic modeling the most widely
used front-end is the log-mel front-end, consisting of melfilterbank energy extraction followed by log compression, where the log compression is used to reduce the dynamic range of filterbank energy. However, there are several issues with the log function.
First, a log has a singularity \cite{flajolet1990singularity} at 0. Common methods to deal with the singularity introduce uncertainity and may have different performance impacts on different signals. Second, the log function uses a lot of its dynamic range on low levels, such as silence, which is likely the least informative part of the signal. Third, the log function is loudness dependent. With different loudness, the log function can
produce different feature values even when the underlying signalcontent (e.g. keywords) is the same, which introduces another factor of variation into training and inference. In their paper, Wang et al \cite{45911} introduce PCEN as an alternative to overcome the above issues with log filterbank. The key ingredient of PCEN is its use of automatic gain control \cite{perez2011automatic}. The PCEN functions are also differentiable and hence they are included as a neural network layer so that it can learn from the attributes of the dataset \cite{45911}. PCEN is well-known to increase the performance of systems that work on keyword spotting tasks, hence we train a model that learns to spot mosquito buzz, using the same far-field detection method as Siri, Alexa, or Google \cite{kim2019query}.

\subsection{Random Masking}
We also randomly warp blocks of frequency channel and time steps. This approach has shown promising results in end-to-end ASR tasks \cite{park2019specaugment}. This step is part of our pre-processing pipeline and we use it as a regularization technique to avoid overfitting.

\subsection{Trainable Kernels}
We move spectrogram generation from CPU to GPU and generate them on the fly in the training loop \cite{cheuk2020nnaudio}. We found an approximately 250\% speedup in generation time moving from an AMD 3.9 GHz 3990X Threadripper CPU to a NVIDIA V-100 GPU.  Additionally, since spectrogram generation has a large number of hyperparameters that become fixed after generation, it becomes difficult to search for optimal parameters during model training. Moving the spectrogram generation as part of the model training process onto the GPU provides several benefits. Hyper-parameters can be searched without running a separate pre-processing pipeline for each iteration.  Additionally, the STFT is implemented as a 1-D convolution, meaning that the kernel is trainable.  Thus, the model can learn a custom Fourier transform suited for our bio-acoustic domain.  Finally, at inference time, there is significant performance improvement and a simpler pipeline that does not require a pre-processing stage.

Figure 4 below shows how an audio signal is processed through our pipeline before going to a deep-learning backbone for prediction.
\vspace{-1em}
\begin{figure}[h]
  \centering
    \includegraphics[width=\linewidth]{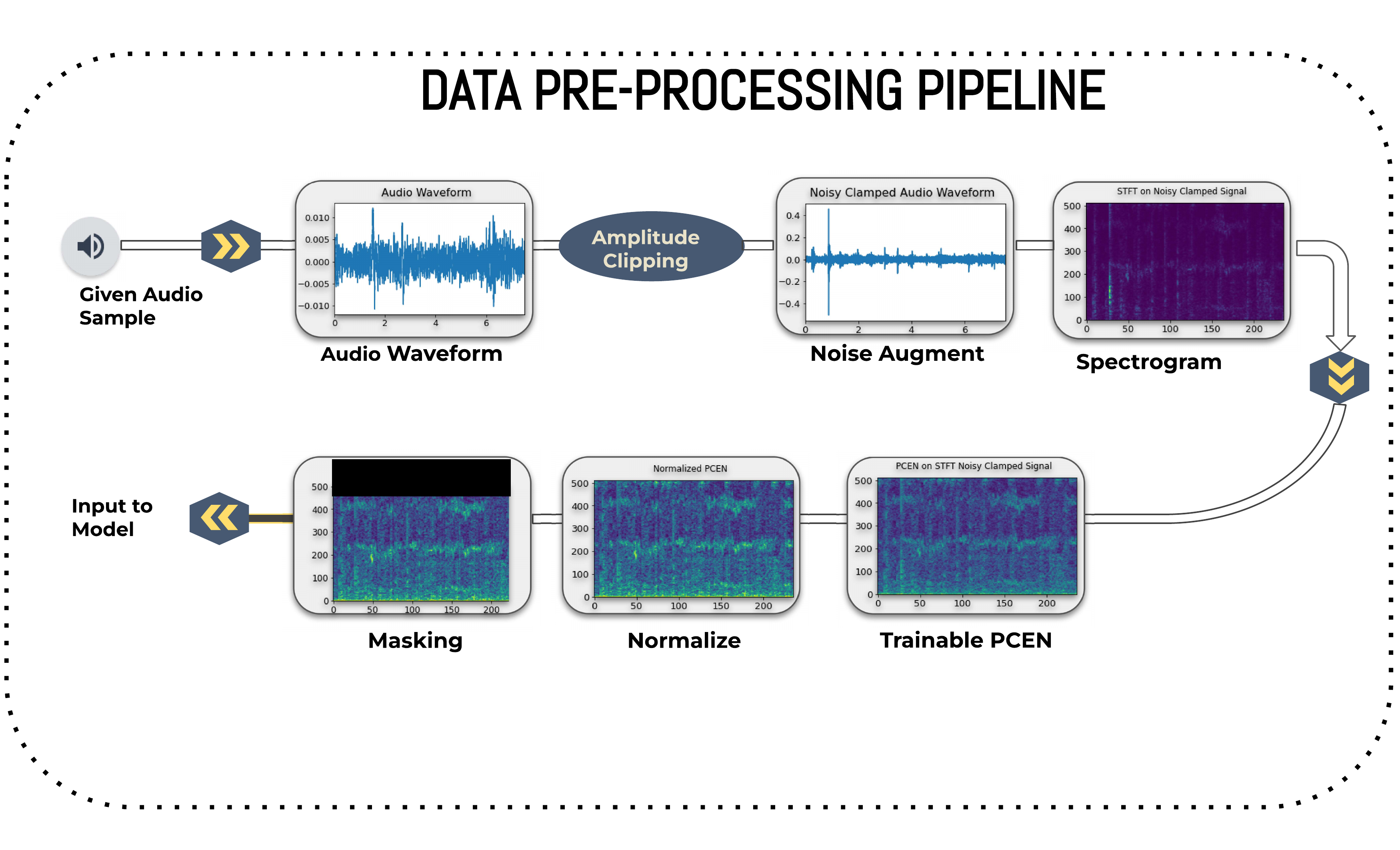}
  \caption{Data Pre-Processing on Raw Audio}
  \Description{This figure denotes how a raw audio file is processed as it moved through the pipeline}
\vspace{-.5em}
\end{figure}

\section{Architecture}
We use Pytorch-based backbones to train our models. Figure 1 above represents an end-to-end pipeline from training to inference. We primarily focus on the Hierarchical Vision Transformer using Shifted Windows (Swin) architecture \cite{liu2021swin} and on CNNs (ConvNeXt) \cite{liu2022convnet}. Our findings indicate that ConvNeXt outperforms Swin on this task. To perform an STFT, we use the default STFT parameters as defined in the baseline configuration\footnote{https://github.com/EIHW/ComParE2022/blob/MOS-C/src/config.py}. 

\section{Results}
Table 1 summarizes the Polyphonic Sound Detection Scores (PSDS) \cite{bilen2020framework} obtained on the development sets by our models.
The training was done on an NVIDIA V-100, 16GB GPU.
We were able to achieve significant improvements on Dev B, which is a more challenging subset that has a lower SNR \cite{Schuller22-TI2}, without causing any reduction to the baseline score on Dev A. Our best-performing model showed an improvement of approximately 2100\% over the published baseline.
The ROC curves for PSDS for ConvNext is depicted in Figure 5.

\begin{table}[htbp]
 \caption{Summary of Dev Results (PSDS)}
  \label{tab:Dev_results}
  \begin{tabular}{cccl}
    \toprule
    Backbone & Dev-Set & Baseline & Best-Score \\
    \midrule
    \textbf{ConvNext} & \textbf{A} & \textbf{.614} & \textbf{.613}\\
    \textbf{ConvNext} & \textbf{B} & \textbf{.034} & \textbf{.758}\\
    Swin & A & .614 & .610\\
    Swin & B & .034 & .040\\
    \bottomrule
  \end{tabular}
\end{table}

\begin{figure}[H]
\begin{tabular}{ll}
\includegraphics[scale=.30]{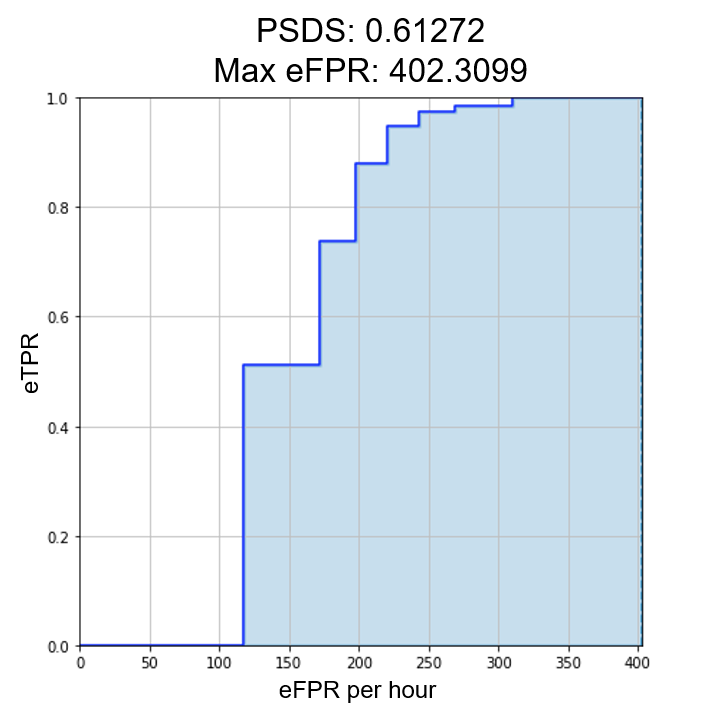}
&
\includegraphics[scale=0.30]{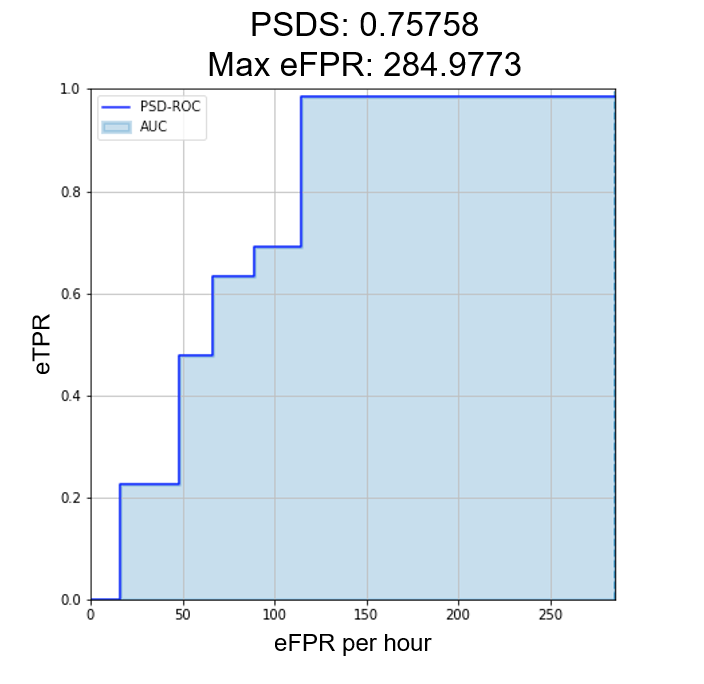}
\end{tabular}
\caption{ConvNext (Left): PSDS Score on Dev A 
(Right): PSDS Score on Dev B}
\label{Fig:Race}
\end{figure} 
\subsection{Test Set Results}
The access to the test set was hidden and submissions were done by leveraging a docker template designed to automatically score the model results over the test data. Out of the allowed five submissions, our best performing model outperformed the published baseline by 212\%.
The table below summarizes the results (PSDS) obtained on the test data.
\vspace{-1em}
\begin{table}[H]
 \caption{Test Set Results (PSDS scores) on a ConvNext Backbone}
 \vspace{-2em}
  \label{tab:commands}
  \begin{tabular}{ccl}
    \toprule
    Data-Set & Baseline & Best-Score\\
    \midrule
    
    Test & .142 & \textbf{.443}\\
    \bottomrule
  \end{tabular}
\end{table}

\section{Experiments and Results: Ablation Study}
To verify the importance of the pre-processing techniques in a quantitative way, we performed an ablation study. We removed each component from our pipeline one at a time and retrained our model with the component missing. Figures 7 below show how the model performed with each component removed. The shorter bars are more important because that means that the model performs significantly worse when that component is missing from the pipeline.  We notice that “random-masking” \cite{park2019specaugment} on spectrograms and “PCEN” are the most significant components of the pipeline; without these components in place, the performance of the model suffers the most. This should come as no surprise, as indicated in the section on Trainable Front End (PCEN); the inclusion of this technique makes our model less vulnerable to the variations in the incoming audio. Random Masking, on the other hand, prevents overfitting.
\begin{figure}[h]
  \centering
    \includegraphics[scale=.25]{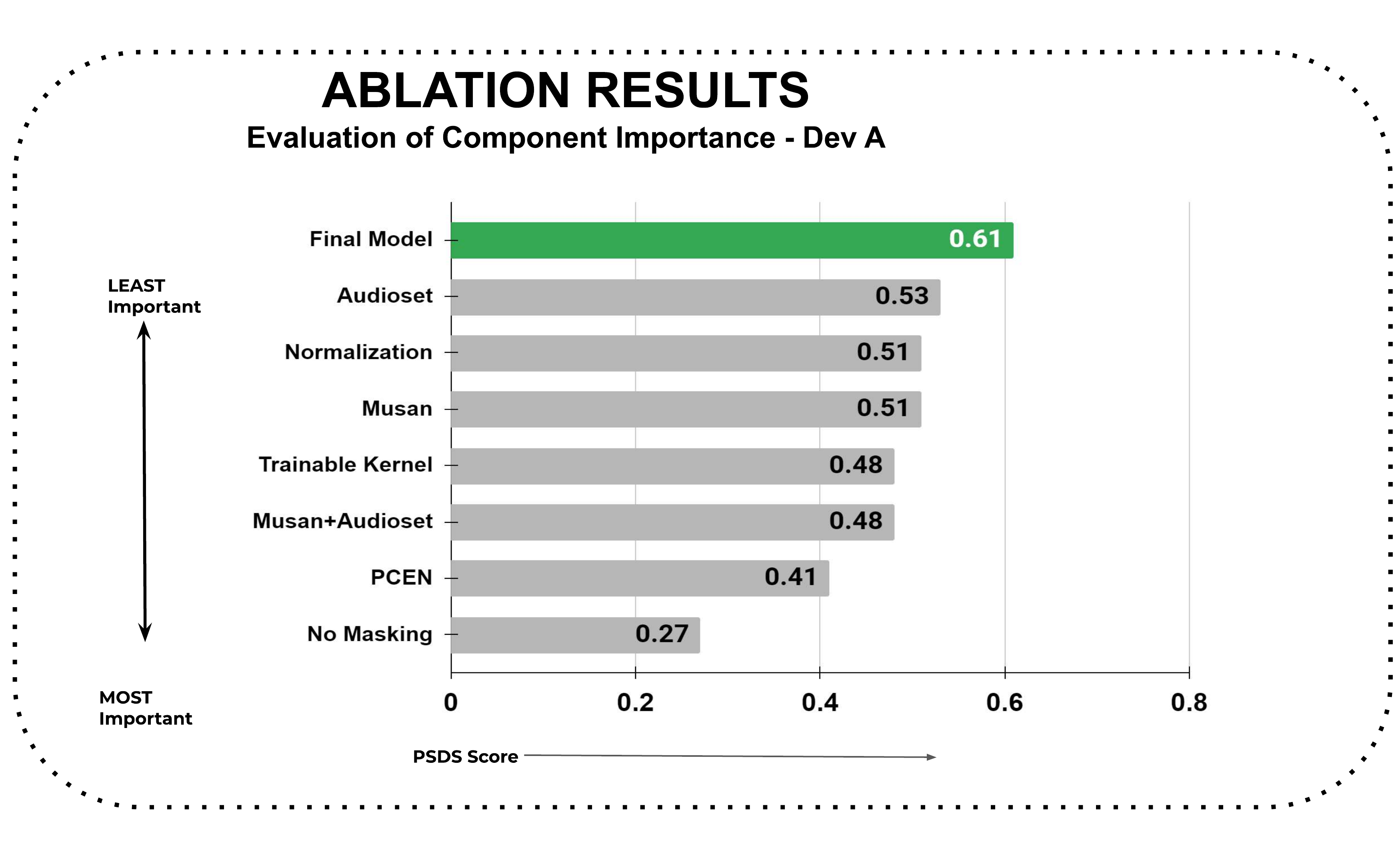}
  \caption{Ablation Study Results - Dev A}
  \Description{This figure denotes how our model performs on Dev A when specific architecture component is removed}

\vspace*{-3.5em}
  \centering
    \includegraphics[scale=.25]{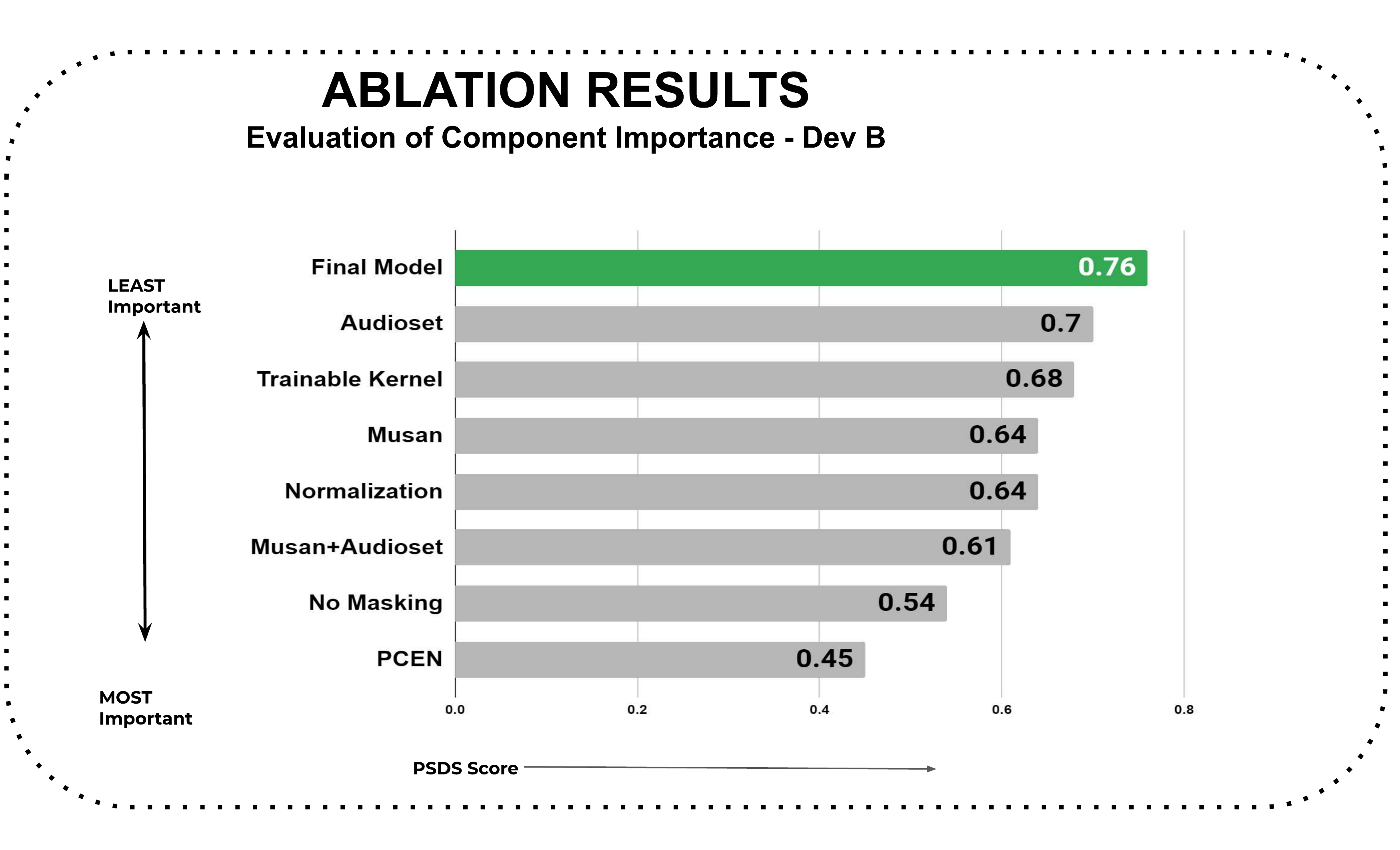}
  \vspace*{-3em}
  \caption{Ablation Study Results}
  \Description{This figure denotes how our model performs on Dev B when specific architecture component is removed}
\vspace*{-2em}
\end{figure}

\section{Conclusion}
Our models outperformed the published baseline on the test set by 212\% and 2100\% on Dev B, which is known to have a very low signal-to-noise ratio (SNR). The improved sensitivity to low SNR events comes without any decrease in performance on Dev A. While detecting the presence of mosquitoes in a real-world setting presents considerable challenges, our results show that with a well-designed pipeline and a judicious choice of neural net architecture, the baseline results can be significantly improved.

\clearpage
\bibliographystyle{ACM-Reference-Format}
\bibliography{Vecnet}


\end{document}